\documentclass[a4paper,fleqn,usenatbib]{mnras}

\def\kms{km~s$^{-1}$}  
\newcommand{\msun}{$M_\odot$}

\newcommand{\lsun}{$L_\odot$}
\usepackage[T1]{fontenc}
\usepackage{ae,aecompl}

\usepackage{graphicx}	% Including figure files
\usepackage{amsmath}	% Advanced maths commands
\usepackage{amssymb}	% Extra maths symbols

\title[The D-type symbiotic EF~Aql]{Optical spectroscopy and X-ray observations of the D-type symbiotic star EF~Aql}

\author[K. A. Stoyanov et al.]{
K. A. Stoyanov,$^{1}$\thanks{E-mail: kstoyanov@astro.bas.bg}
K. I{\l}kiewicz,$^{2}$
G. J. M. Luna,$^{3,4,5}$
J. Miko{\l}ajewska,$^{2}$
K.~Mukai,$^{6,7}$
\newauthor
J.~Mart{\'{\i},}$^{8}$
G.~Latev,$^{1}$
S. Boeva,$^{1}$
and R. K. Zamanov$^{1}$
\\
$^{1}$Institute of Astronomy and National Astronomical Observatory, Bulgarian Academy of Sciences, Tsarigradsko Chaussee 72,\\
BG-1784 Sofia, Bulgaria\\
$^{2}$Nicolaus Copernicus Astronomical Centre, Polish Academy of Sciences, Bartycka 18, P-00716 Warsaw, Poland\\
$^{3}$CONICET-Universidad de Buenos Aires, Instituto de Astronom\'ia y F\'isica del Espacio (IAFE), Av. Inte. G\"uiraldes 2620,\\ 
C1428ZAA, Buenos Aires, Argentina\\
$^{4}$Universidad de Buenos Aires, Facultad de Ciencias Exactas y Naturales, Buenos Aires, Argentina\\
$^{5}$Universidad Nacional de Hurlingham, Av. Gdor. Vergara 2222, Villa Tesei, Buenos Aires, Argentina\\
$^{6}$CRESST and X-ray Astrophysics Laboratory, NASA Goddard Space Flight Center, Greenbelt, MD 20771, USA\\
$^{7}$Department of Physics, University of Maryland, Baltimore County, 1000 Hilltop Circle, Baltimore, MD 21250, USA\\
$^{8}$Departamento de F\'isica (EPSJ), Universidad de Ja\'en, Campus Las Lagunillas,  A3-420, 23071, Ja\'en, Spain
}

\date{Accepted 2020 May 5. Received 2020 May 5; in original form 2020 February 13}

\pubyear{2020}

\begin{document}
\label{firstpage}
\pagerange{\pageref{firstpage}--\pageref{lastpage}}
\maketitle

\begin{abstract}
We performed high-resolution optical spectroscopy and X-ray observations of the recently identified Mira-type symbiotic star EF~Aql. 
Based on high-resolution optical spectroscopy obtained with SALT, we determine the temperature ($\sim $55~000~K) and the luminosity ($\sim$ 5.3~\lsun) of the hot component
in the system. The heliocentric radial velocities of the emission lines in the spectra reveal possible stratification of the chemical elements.
We also estimate the mass-loss rate of the Mira donor star.
Our {\em Swift} observation did not detect EF Aql in X-rays. The upper limit of the X-ray observations is 10$^{-12}$ erg cm$^{-2}$ s$^{-1}$, 
which means that EF~Aql is consistent with the faintest X-ray systems detected so far.
Otherwise we detected it with the UVOT instrument 
with an average UVM2 magnitude of 14.05. During the exposure, EF~Aql became approximately 0.2 UVM2 magnitudes fainter.
The periodogram analysis of the V-band data reveals an improved period of  
320.4$\pm$0.3~d caused by the pulsations of the Mira-type donor star.\\
The spectra are available upon request from the authors. 
\end{abstract}


%%%%%%%%%%%%%%%%%%%%%%%%%%%%%%%%%%%%%%%%%%%%%%%%%%

%%%%%%%%%%%%%%%%% BODY OF PAPER %%%%%%%%%%%%%%%%%%

\section{Introduction}

EF~Aquilae was identified as a variable star on photographic plates from 
K\"onigstuhl Observatory \citep{1925AN....225..385R}. %(Reinmuth 1925)  
\cite{2005JAVSO..34...28R} %Richwine et al. (2005) 
have examined the optical 
photometry for EF~Aql and classify it as a Mira-type variable 
with  a period of 329.4 d and amplitude of variability $> 2.4$ mag in V band.  
\cite{2016PASP..128b4201M} %Margon et al. (2016) 
reported that it has bright UV flux, prominent Balmer emission lines, and 
$[O \: III]\: \lambda 5007$ emission. Thus  EF Aql is classified as a symbiotic star, belonging to the symbiotic Mira subgroup.  
\cite{2017AN....338..680Z} %Zamanov et al. (2017) 
showed that EF~Aql 
displays flickering (stochastic variability with amplitude $\sim 0.2$ magnitudes on a time scale of $\sim 20$ minutes) in B and V bands, 
which is a rarely detectable phenomena among the  symbiotic stars \citep{2001MNRAS.326..553S}.\\%Sokoloski, Bildsten \& Ho 2001 

Symbiotic stars are  long-period interacting binaries, consisting of an yellow or red giant 
transferring mass to a hot compact object. 
According to the classification of \cite{1975MNRAS.171..171W}, %Webster \& Allen (1975) 
the symbiotic stars are divided into two types --
those with cool stellar continuum (S-type) and those with infrared excess from dust (D-type). The D-type symbiotics
have near-IR variations that are caused by the presence of a thermally pulsating Mira-type variable, while in S-type symbiotic stars a normal red giant is present.
\cite{1982ASSL...95...27A} %Allen (1982) 
introduced D'-type symbiotic stars, in which dust is present and a cool giant of spectral type F or G is the donor.
The orbital periods of the symbiotic stars are in the range of 100 days to more than 100 years \citep{2013AcA....63..405G}. %Gromadzki, Miko{\l}ajewska \& Soszy{\'n}ski 2013. 
The D-type symbiotics have longer orbital periods \citep{2013ApJ...770...28H}, %Hinkle et al. 2013 
because the orbits need to be wide enough to harbour a Mira variable with a diameter of several AU 
and a mass of $\sim$ 1 -- 3~\msun\ \citep{1993ApJ...413..641V}. %Vassiliadis \& Wood 1993
In most symbiotic stars, the secondary is a hot and luminous accreting white dwarf \citep{2012BaltA..21....5M}. %Miko{\l}ajewska 2012
In a few cases has the secondary been shown to be 
a neutron star \citep{1977ApJ...211..866D, 2015AstL...41..114K}. %Davidsen, Malina \& Bowyer 1977; Kuranov \& Postnov 2015
The whole symbiotic system is surrounded by a circumstellar nebula formed by the matter lost from the components.

\begin{table}
 \begin{center}
  \caption{UBVR$_C$I$_C$ photometry of EF~Aql obtained on  2019-07-06.}
  \begin{tabular}{llllllll}
       date            & band &    mag            &   \\
\hline       
    2019-07-06 01:49   &  U   &  $14.46 \pm 0.08$ &   \\
    2019-07-06 01:52   &  B   &  $15.41 \pm 0.06$ &   \\
    2019-07-06 01:56   &  B   &  $15.53 \pm 0.07$ &   \\
    2019-07-06 01:58   &  V   &  $14.92 \pm 0.06$ &   \\
    2019-07-06 01:59   &  V   &  $14.96 \pm 0.05$ &   \\
    2019-07-06 01:55   &  R   &  $14.06 \pm 0.03$ &   \\      
    2019-07-06 02:00   &  R   &  $13.97 \pm 0.03$ &   \\  
    2019-07-06 01:56   &  I   &  $11.25	\pm 0.03$ &   \\   
    2019-07-06 02:00   &  I   &  $11.28	\pm 0.03$ &   \\   
 \label{tab3}
 \end{tabular}
 \end{center}
\end{table}

\begin{table*}
 \begin{center}
  \caption{The measured EWs, FWHMs and heliocentric velocities of the lines in the spectrum of EF~Aql. Negative values of EWs denote emission lines, while positive values denote absorption lines.}
  \begin{tabular}{l | c c c | c c c | c c c |  l r r }
 line                 & & {\bf 2019-06-07} & & & {\bf 2019-07-09} & & & {\bf 2019-07-14} & & \\
		      & EW & FWHM & RV$_{hel}$  & EW & FWHM & RV$_{hel}$ & EW & FWHM & RV$_{hel}$ &  \\  
		      & [\AA] & [\kms] & [\kms] & [\AA] & [\kms] & [\kms] & [\AA] & [\kms] & [\kms] & \\
\hline
& &          &           &           & &          &           &           &\\ 
H$\alpha$  6562.82   &  -88.40  & 83 & 3.4  & -90.7    & 74  & 2.8  & -88.8   &	74	& 1.0     & \\
H$\beta$   4861.33   &  -11.37  & 85 & 0.8  & -10.4    & 60  & -0.8 & -11.4   &	59	& -3.1    & \\
H$\gamma$  4340.47   &  -2.78   & 58 & -2.1 & -1.51    & 44  & -3.7 & -3.27   &	44	& -3.2    & \\
                     &          &    &      &	      &     &      &	     &		&         & \\
HeI	4471.48      &  -1.46   & 63 & 7.9  & -1.34    & 71  & -2.1 &  --     &  --      &  --     & \\
HeI     4713.14      &  -0.1    & 38 & 12.3 & -0.06    & 37  & 3.5  & -0.08   & 15       &  -0.8   & \\
HeI     4921.93      &  -1.33   & 91 & 16.2 & -2.92    & 152 & 17   & -3.80   & 120      &  15.9   & \\
HeI     5015.68      &  -2.57   & 115& 19.3 & -3.72    & 132 & 12   & -5.94   & 130      & 21      & \\
HeI     5875.64      &  -3.19   & 67 & 2.4  & -3.31    & 53  & -11.9& -2.07   & 39	& -2.9    & \\
HeI     6678.15      &  -0.95   & 48 & 6.3  & -0.30    & 28  & -2.5 & -0.50   & 33	& -2.1    & \\
HeI     7065.19      &  -1.66   & 52 & 1.5  & -0.90    & 32  & -0.8 & -1.40   & 35	& -1.4    & \\
                     & 	       &    &      &	      &     &      &	     &		&         & \\
$[$OIII$]$  4363.21  &  -0.37   & 41 & -6.3 & -0.39    & 34  & -7.8 & -0.52   & 32	&  -6.0   & \\
$[$OIII$]$  4958.92  &  -0.75   & 25 & -12.2& -0.99    & 27  & -10.9& -1.29   & 25	&  -12.0  & \\
$[$OIII$]$  5006.85  &  -1.66   & 25 & -11.9& -1.57    & 26  & -12.0& -2.56   & 28       &  -12.5  & \\
$[$OI$]$    8446.60  &  -2.69   & 60 & 2.2  & --       & --  & --   & -1.32   & 42	&  -2.3   & \\
                     &          &    &      &          &     &      &         &          &         & \\
FeII~38~4549.47      &  -0.92   & 63 & -0.5 & -0.24    & 36  & 3.4  & --      & --       & --      & \\
FeII~38~4583.83      &  -0.71   & 52 & 4.9  & -0.58    & 50  & 0.8  & -0.54   & 49       & 7.0     & \\
FeII~37~4629.34      &  -0.82   & 90 & 6.9  & -0.58    & 110 & 17.2 & -0.89   & 67       & 3.7     & \\
FeII~42~4923.92      &  -1.95   & 79 & 3.4  & -1.18    & 64  & 12.7 & -2.68   & 88       & 8.6     & \\
FeII~42~5018.43      &  -1.84   & 67 & 4.9  & -1.92    & 74  & 8.4  & -2.61   & 69       & 7.3     & \\
                     &          &    &      &          &     &      &         &          &         & \\
$[$NII$]$~6583.29    & -0.41    & 38 & -11.2& -0.31    & 34  & -13.7& -0.27   & 27       & -5.4    & \\ 
                     &          &    &      &          &     &      &         &          &         & \\
NaI~D1~5889.95       & 0.45     & 19 & -15.8& 0.31     & 14  & -16.7& 0.33    & 16       & -12.8   & \\
NaI~D2~5895.92       & --       & -- & --   & 0.29     & 15  & -16.4& 0.24    & 13       & -13.0   & \\
 \label{tab2}
 \end{tabular}
 \end{center}
\end{table*}

\begin{figure*}
\vspace{0.5cm}
  \vspace{5.5cm} 
  \includegraphics{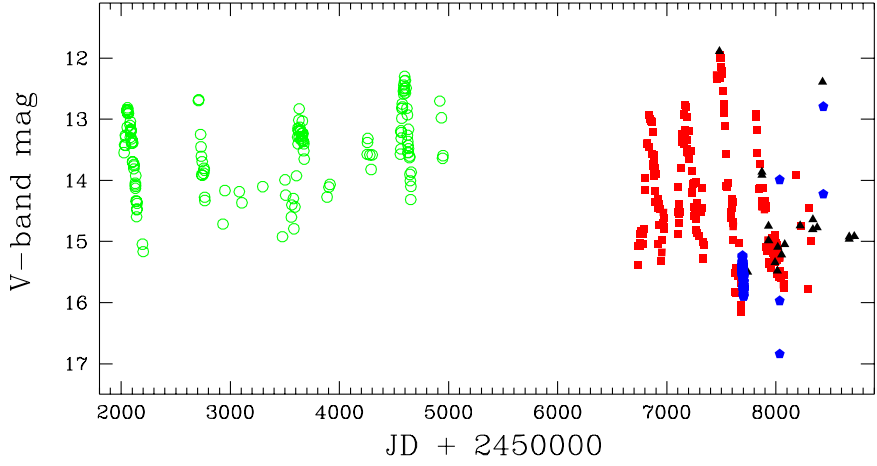} 
  \caption[]{The V-band photometric observations of EF~Aql. The green open circles represent the data from ASAS, the red squares -- the ASAS-SN data,
  the blue pentagons -- the AAVSO data and the black triangles are the observations from Rozhen NAO.}
  \label{phot}
\end{figure*}  

\section{Observations}
\subsection{Photometry}
The V-band light curve of EF~Aql based on data from ASAS \citep{1997AcA....47..467P}, %Pojmanski 1997 
ASAS-SN \citep{2014AAS...22323603S, 2017PASP..129j4502K} %Shappee et al., 2014; Kochanek et al., 2017
and AAVSO International database is shown on Fig.\ref{phot}.
In addition, a UBVR$_C$I$_C$ photometric series of EF~Aql were  obtained in the period 2016 -- 2019 with the 50/70~cm Schmidt telescope of the Rozhen National 
Astronomical Observatory, Bulgaria. Journal of UBVR$_C$I$_C$ observations, obtained on 2019 July 06, is given in Table~\ref{tab3}.

\subsection{SALT observations}
\begin{figure*}   
\vspace{7cm}
  \includegraphics{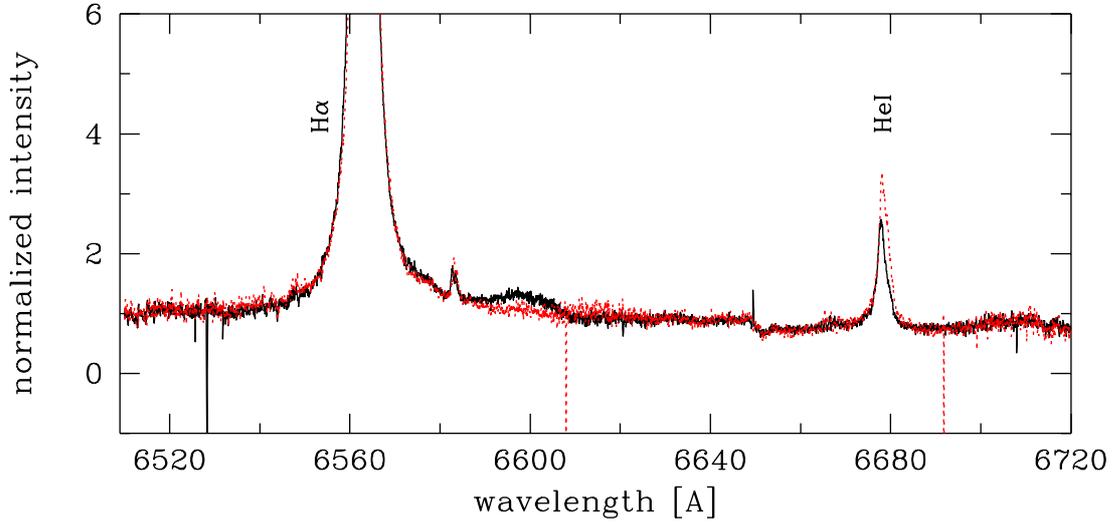} 
  \caption[]{The region around $H\alpha$ in the spectrum of EF~Aql. 
     Black solid line is 2019-06-07, red dashed line -- 2019-07-14.}
  \label{f.Ha} 
\end{figure*}

\begin{figure*}  
\vspace{7cm}
  \includegraphics{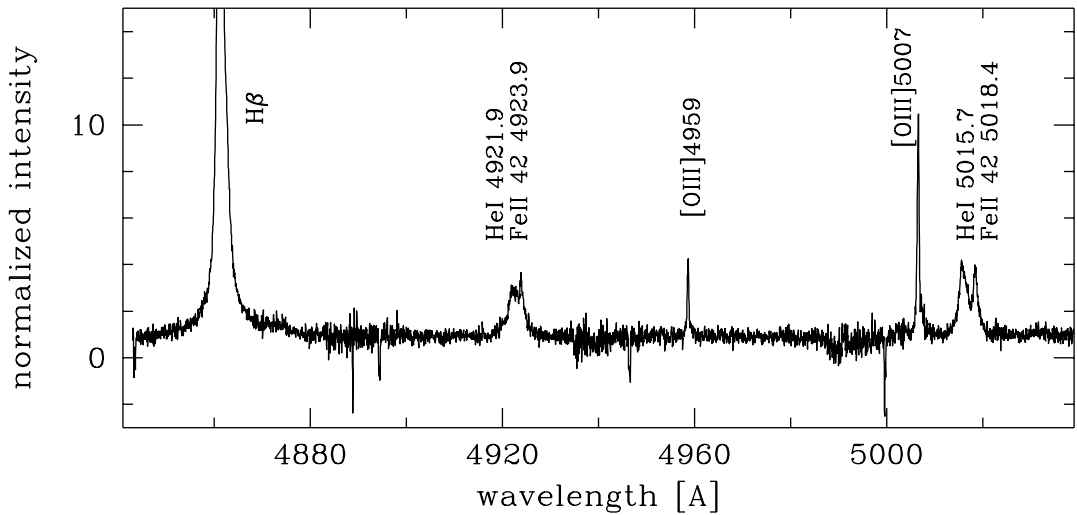}       
  \caption[]{The emission lines around the H$\beta$ line in the spectrum of EF~Aql. }  
  \label{f.Hb} 
\end{figure*}

Spectroscopic observations were carried out on 2019 June 07, 2019 July 09 and 2019 July 14 with the  11m 
Southern African Large Telescope \citep{2006SPIE.6267E..0ZB, 2006MNRAS.372..151O} %Buckley, Swart \& Meiring 2006; O'Donoghue et al. 2006
using the High Resolution Spectrograph \citep{2010SPIE.7735E..4FB, 2012SPIE.8446E..0AB, 2014SPIE.9147E..6TC} %Bramall et al. 2010, 2012; Crause et al. 2014 
The time of observations corresponds to phases $\Phi$=0.61, $\Phi$=0.71 and $\Phi$=0.72 respectively (for details see Section~\ref{per}).
The exposure times are 1900~s.
The spectrograph was used in a medium resolution mode with resolving power R$\sim$40000 
and wavelength coverage of 4000 -- 8800~\AA. 
The initial data reduction was performed using pysalt pipeline \citep{2010SPIE.7737E..25C} %Crawford et al. 2010
which was fallowed by HRS pipeline \citep{2017ASPC..510..480K}, %Kniazev, Gvaramadze \& Berdnikov 2016
based on MIDAS feros \citep{1999ASPC..188..331S} %Stahl, Kaufer \& Tubbesing 1999
and echelle \citep{1992ESOC...41..177B} %Ballester 1992
packages. In Table~\ref{tab2} we present the measured 
equivalent widths (EWs), full widths at half maximum (FWHMs) and heliocentric radial velocities (RV$_{hel}$) of some emission lines in the spectrum of EF~Aql.
The measurements are done by employing a Gaussian fit.
Figures~ \ref{f.Ha} and \ref{f.Hb} show
the regions around the H$\alpha$ and H$\beta$ emission lines.

\subsection{Swift observations}
X-ray and UV observations were obtained using the {\sl Neil Gehrels Swift
observatory\/} (hereafter {\em Swift}) on 2019-09-12 (ObsID: 00011552001). The X-Ray Telescope (XRT)
was operated in photon counting mode, while the UltraViolet and optical
Telescope (UVOT) was operated in imaging mode using UVM2 filter, centered
approximately at 2200~\AA, and both achieved a total exposure of approximately
3.8~ksec.

\section{Results}
\subsection{Period of pulsations}
\label{per}
\begin{figure}   
  \vspace{5.5cm} 
  \includegraphics{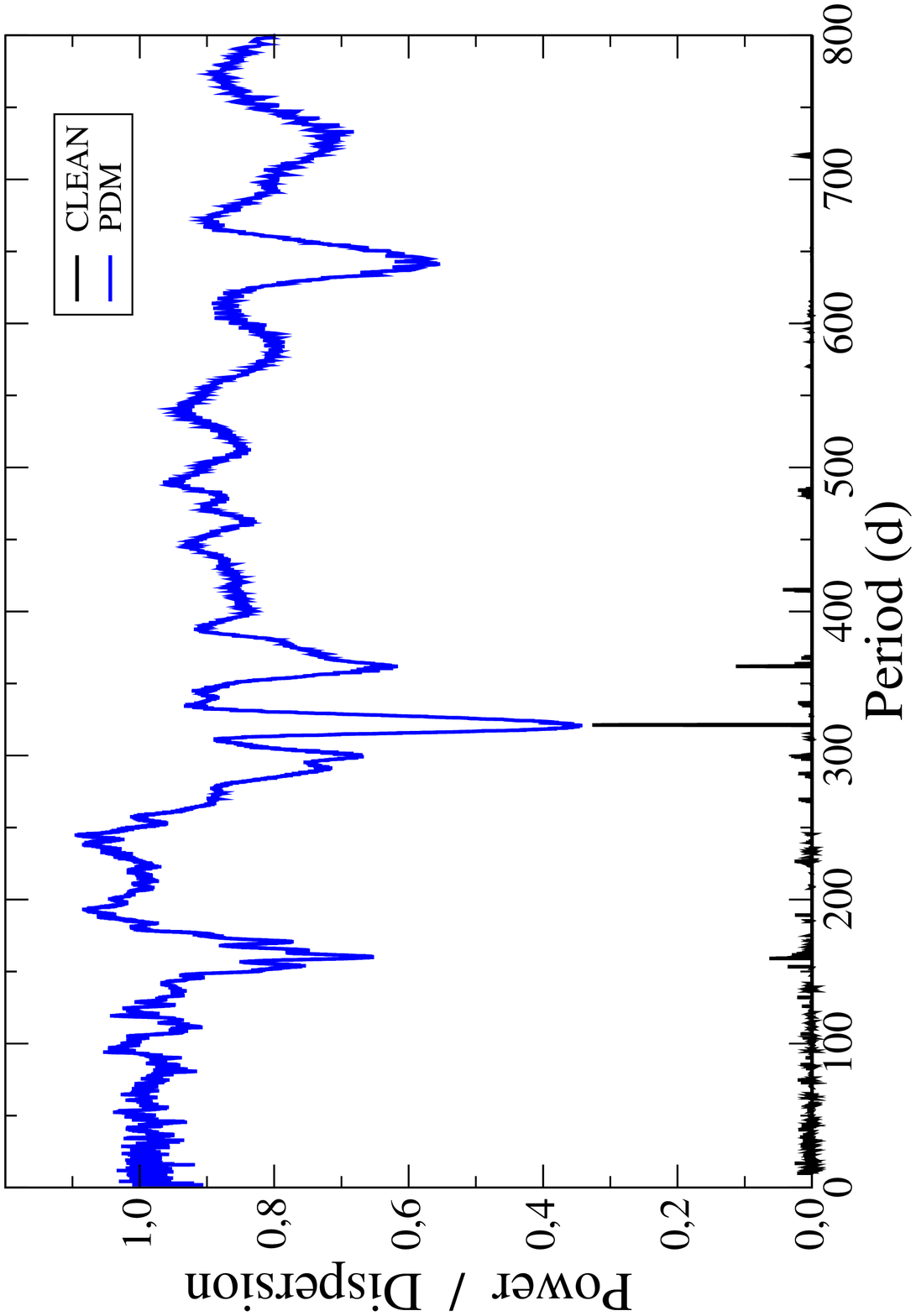} 
  \caption[]{Periodograms for the V-band magnitude of EF~Aql computed using the CLEAN and PDM algorithms. The period of pulsations of the Mira is detected as the most significant one.}
  \label{period}
\end{figure}

Using the photometric data, we performed periodogram analysis in order to confirm and improve the detected period of 
329.4~d, that is associated with the pulsations of the Mira-type donor star \citep{2005JAVSO..34...28R}. %Richwine et al. 2005 
Using the CLEAN algorithm \citep{1987AJ.....93..968R}, %Roberts, Lehar \& Dreher 1987
PDM method \citep{1978ApJ...224..953S} %Stellingwerf 1978 
and Fourier transform we obtained periods of 321.2~$\pm$~0.1~d, 320.8~$\pm$~0.1~d and 320.4~$\pm$~0.3~d respectively.
We present the peridograms of the CLEAN and PDM methods on Fig.\ref{period}. The period and the 
amplitude of the pulsations ($\sim$ 4~mag, see Fig.\ref{phot}) are typical for the Mira-type variables \citep{2003MNRAS.342...86W}. %Whitelock et al. 2003
For the purposes of our study, we use the following ephemeris:

\begin{equation}
 JD_{max} = (2458127.3 \pm 1.6) + (320.4) \pm 0.3 \times E
 \label{eq}
\end{equation}

\begin{figure*} 
\vspace{0.5cm}
  \vspace{5.5cm} 
  \includegraphics{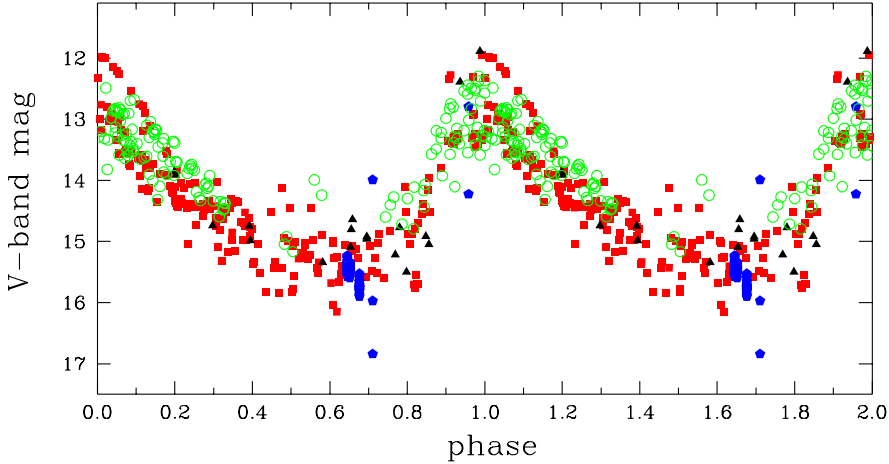} 
  \caption[]{The V-band photometric observations of EF~Aql folded with the period of the pulsations. The symbols are the same as on Fig.\ref{phot}.}
  \label{phase}
\end{figure*}

We present the V-band light curve of EF~Aql folded with the period of 320.4~d on Fig.\ref{phase}.

\begin{figure}
\vspace{6cm}
  \includegraphics{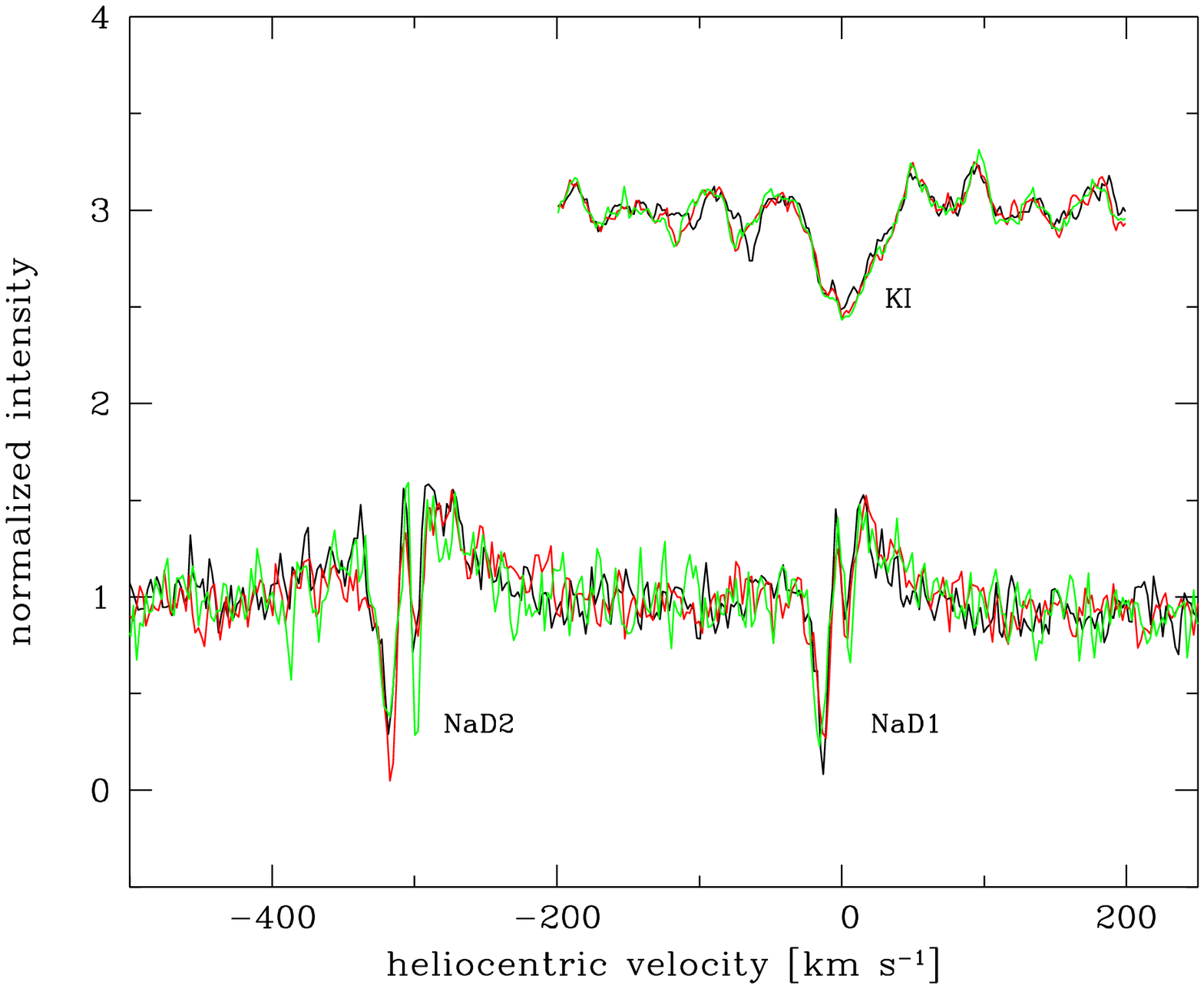} 
  \caption[]{The spectra of EF~Aql in NaD1 and KI7699 lines.
  Black solid line is 2019-06-07, red dashed line --  2019-07-09, green --   2019-07-14.
  Sharp interstellar absorptions are visible in NaD1 and NaD2 lines. 
  No interstellar component is visible in KI~7699.} 
  \label{f.NaD.K}      
\end{figure}   

\subsection{Distance and interstellar reddening}
\label{distance}
As \cite{2003A&A...403..943L} %Le Bertre et al. (2003) 
classified EF~Aql as an O-rich star, we could use the period -- luminosity relations 
for Miras for estimation of the distance to the object \citep{2008MNRAS.386..313W}: %Whitelock, Feast \& Van Leeuwen 2008
\begin{equation}
 M_K = \rho[logP - 2.38] + \delta
 \label{pl}
\end{equation}
where $\rho$ = -3.51 $\pm$ 0.20 and $\delta$ = -7.25 $\pm$ 0.07 \citep{2009AcA....59..169G}. %Gromadzki et al. 2009 
We derive mean K-band magnitude 
from 2MASS All Sky Catalog and DENIS database K=4.78 $\pm$ 0.58 and
absolute K-band magnitude, M$_K$ = -7.69 obtained using Eq.~\ref{pl}.
In \cite{2017AN....338..680Z} %Zamanov et al. (2017) 
we assumed interstellar extinction E(B-V)=0.45, which seems very overestimated.
The NASA/IPAC  IRSA (Galactic Reddening and Extinction Calculator) gives 
low extinction through our galaxy in the direction of EF~Aql,  $E(B-V) < 0.17$,
so the extinction in K-band should be insignificant.  
From these magnitudes we estimate the distance to EF~Aql to be d = 3.1~kpc.\\

In order to estimate the interstellar plus circumstellar extinction, we compare the observed and intrinsic {\it J-K} colours. 
Using the period -- colour relation by \cite{2000MNRAS.319..728W} %Whitelock, Marang \& Feast (2000) 
for O-rich Miras:
\begin{equation}
(J - K)_0 = 0.71 logP - 0.39.
 \label{pc}
\end{equation}

The 2MASS All Sky Catalog and DENIS database give observed {\it J-K} colours 1.61 and 1.81 respectively, which corresponds to E(J-K) = 0.32 $\pm$ 0.10.\\
In our spectra we do not detect DIB 6613 \AA. No interstellar component is visible in KI~7699 line (see Fig.\ref{f.NaD.K}).
The NaI~D lines shows complex profiles with a moderately
broad emission and at least two narrow absorption components. The stronger blue
absorption component may be of local origin, e.g. circumstellar
material, while the red component is of interstellar origin.
The EW of the Na~D1 line is in the range 0.31~\AA\ -- 0.45~\AA\, which corresponds to E(B-V)=0.12 -- 0.25 \citep{1997A&A...318..269M}. %Munari \& Zwitter 1997
It is possible that E(J-K) represents the total (circumstellar and interstellar) extinction and E(B-V) marks only the interstellar extinction.
This would indicate significant circumstellar reddening, as expected for a D-type symbiotic systems \citep{2009AcA....59..169G}. %Gromadzki et al. 2009

\subsection{Temperature and luminosity of the hot component}
The minimum temperature is set by the maximum ionization potential observed in the 
spectrum which in our case is 35.12~eV corresponding to the [OIII] lines \citep{1997iagn.book.....P}. %Peterson 1997
This gives a temperature T$_{hot}$ $\gtrsim$ 35~000~K.
The lack of any traces of HeII lines and the presence of strong HeI lines means that T$_{hot}$ $\lesssim$ 60~000~K.\\

Using the photometric observations listed in Table~\ref{tab3}, we estimate:

\begin{equation}
F_{cont} (R) \sim F_{cont} (6580~$\AA$) \sim 5.6~\times~10^{-15}~erg~s^{-1}~cm^{-2}
 \label{flux}
\end{equation}

This in combination with the measured EW of the H$\alpha$ line (see Table~\ref{tab2}) gives total flux in the H$\alpha$ line:\\ 
F(H$\alpha$)~$\sim$~9~$\times$~10$^{-13}$~erg~s$^{-1}$~cm$^{-2}$.\\

Following the same pattern for the V-band magnitude and the HeI~5876~\AA\ line, and the B-band magnitude and the H$\beta$ lines, the total fluxes in the lines are:\\ 
F(HeI 5876)~$\sim$~4~$\times$~10$^{-14}$~erg~s$^{-1}$~cm$^{-2}$.\\
F(H$\beta$)~$\sim$~1.2~$\times$~10$^{-13}$~erg~s$^{-1}$~cm$^{-2}$.\\

Assuming photionization, case B recombination \citep{1975MNRAS.171..395N} %Netzer 1975 
and blackbody ionizing 
source/hot component, the ratio F(HeI 5876)/F(H$\beta$) indicates 
T$_{hot}$ $\sim $55~000~K, and
the total H$\beta$ and HeI 5876 fluxes imply a hot component 
luminosity L$_{hot}$ $\sim$ 5.3~\lsun\ using d = 3.1~kpc.

\subsection{Mass-loss rate}
The Mira pulsations play a crucial role in the mass-loss process because they raise the dense material from the atmosphere to distances
where the temperature is too low to form dust grains. This is not a well-understood process on the case of the O-rich Miras.  It is believed that a combination of the stellar pulsation 
and radiation pressure on dust is not enough to drive the mass loss \citep{2006A&A...460L...9W} %Woitke 2006
and the proper mechanism is a subject of debates.\\

To estimate the mass loss for EF~Aql, we use the correlation between the mass-loss rate and the K-[12] colour for O-rich Mira variables.
The K and [12] fluxes originate from the star and the dust shell respectively, so larger K-[12] colour means thicker shell. Therefore, the K-[12] colour 
provides a useful tool for mass-loss rate estimation. Using the mean K-band magnitude (see Sect.~\ref{distance}) and the flux at 12~$\mu$m provided by IRAS database 
\citep{1984ApJ...278L...1N}, %Neugebauer et al. 1984 
we estimate K-[12] = 2.89. Using Fig.21 from  \cite{1994MNRAS.267..711W}, %Whitelock et al. (1994) 
we estimate the mass-loss rate for EF~Aql to be $\sim$ 2.5~$\times$~10$^{-7}$~\msun~yr$^{-1}$. 
The mass-loss rates of the single O-rich Miras are in the range from 10$^{-7}$~\msun~yr$^{-1}$ to 10$^{-5}$~\msun~yr$^{-1}$ \citep{1994MNRAS.267..711W}. %Whitelock et al. 1994). 
The Miras in the symbiotic systems have an average
mass-loss rate $\sim$3.2~$\times$~10$^{-6}$~\msun~yr$^{-1}$ \citep{2009AcA....59..169G}. %Gromadzki et al. 2009 
The measured low value for EF~Aql is in agreement with our considerations for a low-luminosity system 
(see the Discussion section).

\subsection{X-ray and UV emission}
Our {\em Swift} observation did not detect EF Aql in X-rays, with a upper limit of 0.003 c s$^{-1}$. Assuming an absorbing column of 10$^{23}$ cm$^{-2}$ and a plasma temperature 
of 10~keV (typical of $\delta$ component in symbiotics), this upper limit on the count rate translates into an unabsorbed flux of 10$^{-12}$ erg cm$^{-2}$ s$^{-1}$ 
and into a luminosity of about half a solar luminosity. This upper limit is consistent with the faintest $\delta$ components detected so far (the component from the boundary layer between the accretion 
disc and the white dwarf; \citet{2013A&A...559A...6L}). %Luna et al. 2013
In turn, we detected EF~Aql with the UVOT instrument 
onboard Swift, with an average UVM2 magnitude of 14.05 (uncorrected for reddening). During the 25 ks that {\em Swift} was pointing to EF~Aql, 
we observed that it got approximately 0.2 UVM2 magnitudes fainter. We can only speculate about this behavior, which could be due to variable 
absorption and/or flickering from the accretion disk.

\begin{figure}
\vspace{5cm}
  \includegraphics{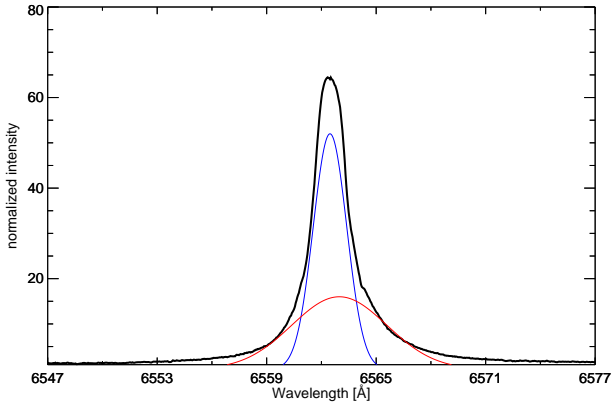} 
  \caption[]{The broad and narrow components of the H$\alpha$ line fitted with two Gaussians with FWHMs $\sim$ 1.8~\AA\ and 6.0~\AA\ respectively.}
  \label{gauss}      
\end{figure}  

\begin{figure}
\vspace{12cm}
  \includegraphics{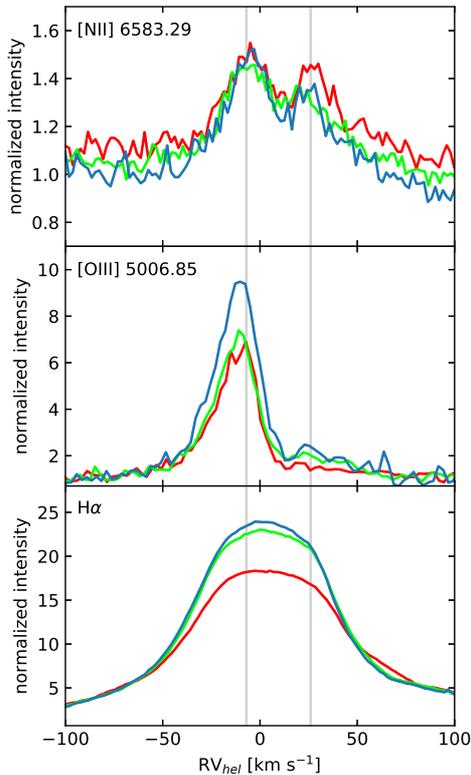} 
  \caption[]{Comparison of emission lines
profiles of EF Aql. The gray lines indicate the positions of two maxima in
the [NII] 6583.29 line.  Red line: SALT 2019.06.06 spectrum, green line:
SALT 2019.07.09 spectrum, blue line: SALT 2019.07.13 spectrum.}
  \label{lines_comp}      
\end{figure}   

\section{Discussion}   
The Balmer lines show complex profiles with a 
relatively broad and a narrow component (Fig.~\ref{gauss}). The same structure is present 
in the strongest HeI lines (4471~\AA, 4713~\AA, 5876~\AA, 6678~\AA\ and 7065~\AA) while it is less 
obvious in the fainter lines. In the case of HeI~4922 and HeI~5016 the nearby 
FeII lines may hide the fine structure of the HeI lines (Fig.~\ref{f.Hb}). No HeII lines are detected in the spectra.\\

The [NII]~6583 line seems to have double-peaked structure with the red component ~50 -- 60\% of the blue one (Fig.~\ref{lines_comp})
The same profile appears 
on all spectra. In the 
case of [OIII] lines, only the blue component is visible. The line 
profile is very similar to that in planetary nebulae \citep{1989PASP..101..966L} %Lutz et al. 1989 
and it may be originate in a faint fossil nebula ejected by the present white dwarf component during its AGB stage \citep{1993A&A...277..195M}.\\ %Munari \& Patat 1993)

It is also interesting that the H$\alpha$ profile appears to have flat top, and, in principle, the
narrow component may consist of two narrow lines with velocity
separation similar to that found in [NII] profiles.\\ 

The temperature of the hot components in symbiotics is in the range 35~000~K -- 500~000~K and the luminosity is in the range 0.3~\lsun -- 37~000~\lsun 
\citep{1991A&A...248..458M, 1994A&A...282..586M, 1997A&A...327..191M, 2007ApJ...661.1105O}. %M\"{u}rset et al. 1991; Murset \& Nussbaumer 1994; Miko{\l}ajewska, Acker \& Stenholm 1997; Orio et al. 2007).
\cite{1997A&A...327..191M} %Miko{\l}ajewska, Acker \& Stenholm (1997) 
found that the temperature of the hot component correlates with the Raman scattered $\lambda$ 6825 line. The 6825~\AA\ emission is produced due to 
Raman scattering of O~VI~$\lambda$1032 by hydrogen atoms \citep{1989A&A...211L..31S}. %Schmid 1989 
The lack of $\lambda$ 6825 emission in EF~Aql supports our result for a hot component with low luminosity.\\

The typical red giant in symbiotic system have a 
mass-loss rate of $\sim$ 10$^{-7}$~\msun~yr$^{-1}$, which is enough to power a luminosity of 10 -- 100~\lsun. In systems with hydrogen shell burning on the surface of the white dwarf, 
even an accretion rate of $\sim$ 10$^{-8}$~\msun~yr$^{-1}$ can produce that amount of luminosity. The Mira-type variables have a mass loss rate of $\sim$ 10$^{-5}$~\msun~yr$^{-1}$ --  
10$^{-7}$~\msun~yr$^{-1}$ \citep{1994MNRAS.267..711W}, %Whitelock et al. 1994 
but the larger distance between the components in the D-type symbiotic stars results similar luminosities. In the systems without hydrogen 
shell burning on the surface of the white dwarf (the case of EF~Aql), the luminosity of the hot component is more or less equal to that generated by the accretion 
\citep{1978ApJ...222..604P}.\\ %Paczynski \& Zytkow 1978)

The obtained parameters set up EF~Aql among the systems with lowest temperature and luminosity of the hot component and the most probable reason is the low accretion rate.\\

The AGB stage is the final stage of the evolution of the low- and intermediate-mass stars before they turn into white dwarfs. The evolution during this stage is driven by a large range of 
complex processes, including the stellar pulsations and the mass loss. The AGB stars can reach mass-loss rates so high that mass loss process can determines the evolutionary time scales, instead of
nuclear burning time scale. Large-scale convective flows bring newly-formed chemical elements from the stellar interior to the surface and, boost by 
the stellar pulsations, they trigger shock waves in the extended stellar atmosphere, where they formed massive outflows of gas and dust. These outflows not only enrich the intestellar medium, 
but also play a cricial role in the
evolution of the cool giant into a white dwarf \citep{2018A&ARv..26....1H}. %H{\"o}fner \& Olofsson 2018
The progenitors of the short-period Miras have a mass of 1.0 -- 1.1~\msun\ \citep{1994ApJ...422..102J}, %Jura 1994
while the Miras with periods in the range 400 -- 600~d have progenitors with masses $\geq$ 2~\msun. 
The lifetime of a star with initial mass 1 -- 2~\msun\ during the AGB stage is in the range 1 -- 3$\times$10$^6$~yr \citep{2014ApJ...782...17K}. %Kalirai, Marigo \& Tremblay 2014
For a Mira with period $\sim$ 300~d, we should expect an increase of the period and the mass loss \citep{2019MNRAS.482..929T}. %Trabucchi et al. 2019)
Following these considerations, for the future evolution of EF~Aql we are expecting an increase of the accretion rate onto the white dwarf followed by an increase of the X-ray 
activity but the evolution of EF~Aql in general should be consider in terms of the binary evolution.\\

EF~Aql appears to be powered by
accretion onto a white dwarf without shell nuclear burning on the surface of the white dwarf. The presence of optical
flickering \citep{2017AN....338..680Z} %Zamanov et al. 2017 
and the modest hot component luminosity (this work)
are both consistent with this interpretation. While we did not detect EF Aql in
X-rays, particularly X-rays of the $\delta$-type \citep{2013A&A...559A...6L}, %Luna et al. 2013
our upper limit is still consistent with lower-luminosity end of
accretion-powered symbiotics.\\

The measured heliocentric radial velocities (see Table~\ref{tab2}) point for a possible ionization potential dependent stratification in EF~Aql. Similar effect is detected 
in the D-type symbiotic RR~Tel \citep{2004ASPC..318..363K} %Kotnik-Karuza et al. 2004 
and in some S-type symbiotic stars \citep{2010A&A...512A..80F}. %Friedjung et al. 2010 
A velocity stratification of various spectral lines is also detected in
some Mira-type variables \citep{1992A&A...253..203S}. %Scholz 1992
Single Mira-type variables can usually produce many strong emission lines in their spectra \citep{1960stat.book..509M, 1984ApJ...286..337F}, %Merrill 1960; Fox, Wood \& Dopita 1984 
e. g. the FeII lines. On the other hand, the broad Balmer 
lines and the forbidden lines of the oxygen originates from the circumbinary material.\\

\section{Conclusions}
We obtained high-resolution optical spectroscopy and X-ray observations of the D-type symbiotic star EF~Aql.
Using the spectra and additional photometry, we estimated the temperature (T$_{hot}$ $\sim $55~000~K) and the luminosity (L$_{hot}$ $\sim$ 5.3~\lsun) 
of the hot component in the system. The measured parameters
of the emission lines in the spectra, reveal possible ionization potential dependent stratification.
We performed periodogram analysis on the available V-band photometric data and found an improved period of 
320.4$\pm$0.3~d, which is the period of pulsations of the Mira-type donor star.\\

The {\em Swift} observation did not detect EF~Aql in X-rays. 
The upper limit of the X-ray observations is 0.003 c s$^{-1}$ which, for a
$\delta$-type spectrum, corresponds into an unabsorbed flux of 10$^{-12}$ erg cm$^{-2}$ s$^{-1}$. 
This means that EF~Aql may well be comparable to the X-ray faintest
$\delta$-type symbiotics detected so far.
However, we detected EF~Aql in the UV 
with an average UVM2 magnitude of 14.05.
Based on IR data, we estimate the distance to EF~Aql to be $\sim$ 3.1~kpc and a mass-loss rate of the Mira donor of $\sim$ 2.5~$\times$~10$^{-7}$~\msun~yr$^{-1}$. \\
The optical and X-ray observations point that EF~Aql is an accretion-powered symbiotic star without shell burning.
The results are proof that the D-type symbiotic stars deserve more attention and observations,
which can help to understand the binary companions to asymptotic giant branch (AGB) stars and their evolution.

\section*{Acknowledgements}
We  thank  the  anonymous  referee  for  the  comments  which  improved the paper.
This work is supported by the grant K$\Pi$-06-H28/2 08.12.2018 
(Bulgarian National Science Fund). The paper is based on spectroscopic observations made with the Southern 
African Large Telescope (SALT) under programme
2018-1-MLT-005 (PI: J. Miko{\l}ajewska). Polish participation in SALT is 
funded by grant No. MNiSW DIR/WK/2016/07.
This research has been partly founded by the National Science Centre, 
Poland, through grant OPUS 2017/27/B/ST9/01940 to J. Miko{\l}ajewska.
GJML is a member of the CIC-CONICET (Argentina) and acknowledge support from grant ANPCYT-PICT 0901/2017.
J.~Mart{\'{\i} acknowledges support by Agencia Estatal de Investigaci\'on grant 
AYA2016-76012-C3-3-P 
from the Spanish Ministerio de Econom\'{\i}a y Competitividad (MINECO), and 
by Consejer\'{\i}a de Econom\'{\i}a, Innovaci\'on, Ciencia y Empleo of Junta de
Andaluc\'{\i}a 
under research group FQM-322, as well as FEDER funds.\\ 
We thank Brad Cenko, the PI of the {\sl Neil Gehrels Swift\/} observatory, for a general allocation of observing time.
This research has made use of the NASA/ IPAC Infrared Science Archive, which is operated by the Jet Propulsion Laboratory, 
California Institute of Technology, under contract with the National Aeronautics and Space Administration and the XRT Data 
Analysis Software (XRTDAS) developed under the responsibility of the ASI Science Data Center (ASDC), Italy.
The authors acknowledge the variable star observations from the AAVSO International Database contributed by observers worldwide and used in this research.
ASAS-SN is supported by the Gordon and Betty Moore
Foundation through grant GBMF5490 to the Ohio State
University and NSF grant AST-1515927. Development of
ASAS-SN has been supported by NSF grant AST-0908816,
the Mt. Cuba Astronomical Foundation, the Center for Cosmology and AstroParticle Physics at the Ohio State University, the Chinese Academy of Sciences South America Center
for Astronomy (CAS-SACA), the Villum Foundation, and
George Skestos. 

%%%%%%%%%%%%%%%%%%%%%%%%%%%%%%%%%%%%%%%%%%%%%%%%%%



%\bibliographystyle{mnras}


\bibliographystyle{mnras}
\bibliography{ref.bib}

\begin{thebibliography}{}
\makeatletter
\relax
\def\mn@urlcharsother{\let\do\@makeother \do\$\do\&\do\#\do\^\do\_\do\%\do\~}
\def\mn@doi{\begingroup\mn@urlcharsother \@ifnextchar [ {\mn@doi@}
  {\mn@doi@[]}}
\def\mn@doi@[#1]#2{\def\@tempa{#1}\ifx\@tempa\@empty \href
  {http://dx.doi.org/#2} {doi:#2}\else \href {http://dx.doi.org/#2} {#1}\fi
  \endgroup}
\def\mn@eprint#1#2{\mn@eprint@#1:#2::\@nil}
\def\mn@eprint@arXiv#1{\href {http://arxiv.org/abs/#1} {{\tt arXiv:#1}}}
\def\mn@eprint@dblp#1{\href {http://dblp.uni-trier.de/rec/bibtex/#1.xml}
  {dblp:#1}}
\def\mn@eprint@#1:#2:#3:#4\@nil{\def\@tempa {#1}\def\@tempb {#2}\def\@tempc
  {#3}\ifx \@tempc \@empty \let \@tempc \@tempb \let \@tempb \@tempa \fi \ifx
  \@tempb \@empty \def\@tempb {arXiv}\fi \@ifundefined
  {mn@eprint@\@tempb}{\@tempb:\@tempc}{\expandafter \expandafter \csname
  mn@eprint@\@tempb\endcsname \expandafter{\@tempc}}}

\bibitem[\protect\citeauthoryear{{Allen}}{{Allen}}{1982}]{1982ASSL...95...27A}
{Allen} D.~A.,  1982, {Infrared studies of symbiotic stars.}.
pp 27--42

\bibitem[\protect\citeauthoryear{{Ballester}}{{Ballester}}{1992}]{1992ESOC...41..177B}
{Ballester} P.,  1992, in European Southern Observatory Conference and Workshop
  Proceedings. p.~177

\bibitem[\protect\citeauthoryear{{Bramall} et~al.,}{{Bramall}
  et~al.}{2010}]{2010SPIE.7735E..4FB}
{Bramall} D.~G.,  et~al., 2010, {The SALT HRS spectrograph: final design,
  instrument capabilities, and operational modes}.
p. 77354F

\bibitem[\protect\citeauthoryear{{Bramall} et~al.,}{{Bramall}
  et~al.}{2012}]{2012SPIE.8446E..0AB}
{Bramall} D.~G.,  et~al., 2012, {The SALT HRS spectrograph: instrument
  integration and laboratory test results}.
p. 84460A

\bibitem[\protect\citeauthoryear{{Buckley}, {Swart}  \& {Meiring}}{{Buckley}
  et~al.}{2006}]{2006SPIE.6267E..0ZB}
{Buckley} D. A.~H.,  {Swart} G.~P.,   {Meiring} J.~G.,  2006, {Completion and
  commissioning of the Southern African Large Telescope}.
p. 62670Z

\bibitem[\protect\citeauthoryear{{Crause} et~al.,}{{Crause}
  et~al.}{2014}]{2014SPIE.9147E..6TC}
{Crause} L.~A.,  et~al., 2014, {Performance of the Southern African Large
  Telescope (SALT) High Resolution Spectrograph (HRS)}.
p. 91476T

\bibitem[\protect\citeauthoryear{{Crawford} et~al.,}{{Crawford}
  et~al.}{2010}]{2010SPIE.7737E..25C}
{Crawford} S.~M.,  et~al., 2010, {PySALT: the SALT science pipeline}.
p. 773725

\bibitem[\protect\citeauthoryear{{Davidsen}, {Malina}  \& {Bowyer}}{{Davidsen}
  et~al.}{1977}]{1977ApJ...211..866D}
{Davidsen} A.,  {Malina} R.,   {Bowyer} S.,  1977, \mn@doi [\apj]
  {10.1086/154996}, \href
  {https://ui.adsabs.harvard.edu/abs/1977ApJ...211..866D} {211, 866}

\bibitem[\protect\citeauthoryear{{Fox}, {Wood}  \& {Dopita}}{{Fox}
  et~al.}{1984}]{1984ApJ...286..337F}
{Fox} M.~W.,  {Wood} P.~R.,   {Dopita} M.~A.,  1984, \mn@doi [\apj]
  {10.1086/162604}, \href
  {https://ui.adsabs.harvard.edu/abs/1984ApJ...286..337F} {286, 337}

\bibitem[\protect\citeauthoryear{{Friedjung}, {Miko{\l}ajewska}, {Zajczyk}  \&
  {Eriksson}}{{Friedjung} et~al.}{2010}]{2010A&A...512A..80F}
{Friedjung} M.,  {Miko{\l}ajewska} J.,  {Zajczyk} A.,   {Eriksson} M.,  2010,
  \mn@doi [\aap] {10.1051/0004-6361/200913438}, \href
  {https://ui.adsabs.harvard.edu/abs/2010A&A...512A..80F} {512, A80}

\bibitem[\protect\citeauthoryear{{Gromadzki}, {Miko{\l}ajewska}, {Whitelock}
  \& {Marang}}{{Gromadzki} et~al.}{2009}]{2009AcA....59..169G}
{Gromadzki} M.,  {Miko{\l}ajewska} J.,  {Whitelock} P.,   {Marang} F.,  2009,
  \actaa, \href {https://ui.adsabs.harvard.edu/abs/2009AcA....59..169G} {59,
  169}

\bibitem[\protect\citeauthoryear{{Gromadzki}, {Miko{\l}ajewska}  \&
  {Soszy{\'n}ski}}{{Gromadzki} et~al.}{2013}]{2013AcA....63..405G}
{Gromadzki} M.,  {Miko{\l}ajewska} J.,   {Soszy{\'n}ski} I.,  2013, \actaa,
  \href {https://ui.adsabs.harvard.edu/abs/2013AcA....63..405G} {63, 405}

\bibitem[\protect\citeauthoryear{{Hinkle}, {Fekel}, {Joyce}  \&
  {Wood}}{{Hinkle} et~al.}{2013}]{2013ApJ...770...28H}
{Hinkle} K.~H.,  {Fekel} F.~C.,  {Joyce} R.~R.,   {Wood} P.,  2013, \mn@doi
  [\apj] {10.1088/0004-637X/770/1/28}, \href
  {https://ui.adsabs.harvard.edu/abs/2013ApJ...770...28H} {770, 28}

\bibitem[\protect\citeauthoryear{{H{\"o}fner} \& {Olofsson}}{{H{\"o}fner} \&
  {Olofsson}}{2018}]{2018A&ARv..26....1H}
{H{\"o}fner} S.,  {Olofsson} H.,  2018, \mn@doi [\aapr]
  {10.1007/s00159-017-0106-5}, \href
  {https://ui.adsabs.harvard.edu/abs/2018A&ARv..26....1H} {26, 1}

\bibitem[\protect\citeauthoryear{{Jura}}{{Jura}}{1994}]{1994ApJ...422..102J}
{Jura} M.,  1994, \mn@doi [\apj] {10.1086/173707}, \href
  {https://ui.adsabs.harvard.edu/abs/1994ApJ...422..102J} {422, 102}

\bibitem[\protect\citeauthoryear{{Kalirai}, {Marigo}  \& {Tremblay}}{{Kalirai}
  et~al.}{2014}]{2014ApJ...782...17K}
{Kalirai} J.~S.,  {Marigo} P.,   {Tremblay} P.-E.,  2014, \mn@doi [\apj]
  {10.1088/0004-637X/782/1/17}, \href
  {https://ui.adsabs.harvard.edu/abs/2014ApJ...782...17K} {782, 17}

\bibitem[\protect\citeauthoryear{{Kniazev}, {Gvaramadze}  \&
  {Berdnikov}}{{Kniazev} et~al.}{2017}]{2017ASPC..510..480K}
{Kniazev} A.~Y.,  {Gvaramadze} V.~V.,   {Berdnikov} L.~N.,  2017, {SALT
  Spectroscopy of Evolved Massive Stars}.
p.~480

\bibitem[\protect\citeauthoryear{{Kochanek} et~al.,}{{Kochanek}
  et~al.}{2017}]{2017PASP..129j4502K}
{Kochanek} C.~S.,  et~al., 2017, \mn@doi [\pasp] {10.1088/1538-3873/aa80d9},
  \href {https://ui.adsabs.harvard.edu/abs/2017PASP..129j4502K} {129, 104502}

\bibitem[\protect\citeauthoryear{{Kotnik-Karuza}, {Friedjung}, {Exter},
  {Keenan}, {Pollacco}  \& {Whitelock}}{{Kotnik-Karuza}
  et~al.}{2004}]{2004ASPC..318..363K}
{Kotnik-Karuza} D.,  {Friedjung} M.,  {Exter} K.,  {Keenan} F.~P.,  {Pollacco}
  D.~L.,   {Whitelock} P.~A.,  2004, {New results about dust in the envelope of
  the symbiotic nova RR Tel}.
pp 363--366

\bibitem[\protect\citeauthoryear{{Kuranov} \& {Postnov}}{{Kuranov} \&
  {Postnov}}{2015}]{2015AstL...41..114K}
{Kuranov} A.~G.,  {Postnov} K.~A.,  2015, \mn@doi [Astronomy Letters]
  {10.1134/S1063773715040064}, \href
  {https://ui.adsabs.harvard.edu/abs/2015AstL...41..114K} {41, 114}

\bibitem[\protect\citeauthoryear{{Le Bertre}, {Tanaka}, {Yamamura}  \&
  {Murakami}}{{Le Bertre} et~al.}{2003}]{2003A&A...403..943L}
{Le Bertre} T.,  {Tanaka} M.,  {Yamamura} I.,   {Murakami} H.,  2003, \mn@doi
  [\aap] {10.1051/0004-6361:20030461}, \href
  {https://ui.adsabs.harvard.edu/abs/2003A&A...403..943L} {403, 943}

\bibitem[\protect\citeauthoryear{{Luna}, {Sokoloski}, {Mukai}  \&
  {Nelson}}{{Luna} et~al.}{2013}]{2013A&A...559A...6L}
{Luna} G.~J.~M.,  {Sokoloski} J.~L.,  {Mukai} K.,   {Nelson} T.,  2013, \mn@doi
  [\aap] {10.1051/0004-6361/201220792}, \href
  {https://ui.adsabs.harvard.edu/abs/2013A&A...559A...6L} {559, A6}

\bibitem[\protect\citeauthoryear{{Lutz}, {Kaler}, {Shaw}, {Schwarz}  \&
  {Aspin}}{{Lutz} et~al.}{1989}]{1989PASP..101..966L}
{Lutz} J.~H.,  {Kaler} J.~B.,  {Shaw} R.~A.,  {Schwarz} H.~E.,   {Aspin} C.,
  1989, \mn@doi [\pasp] {10.1086/132560}, \href
  {https://ui.adsabs.harvard.edu/abs/1989PASP..101..966L} {101, 966}

\bibitem[\protect\citeauthoryear{{Margon}, {Prochaska}, {Tejos}  \&
  {Monroe}}{{Margon} et~al.}{2016}]{2016PASP..128b4201M}
{Margon} B.,  {Prochaska} J.~X.,  {Tejos} N.,   {Monroe} T.,  2016, \mn@doi
  [\pasp] {10.1088/1538-3873/128/960/024201}, \href
  {https://ui.adsabs.harvard.edu/abs/2016PASP..128b4201M} {128, 024201}

\bibitem[\protect\citeauthoryear{{Merrill}}{{Merrill}}{1960}]{1960stat.book..509M}
{Merrill} P.~W.,  1960, {Spectra of Long-Period Variables}.
p.~509

\bibitem[\protect\citeauthoryear{{Miko{\l}ajewska}}{{Miko{\l}ajewska}}{2012}]{2012BaltA..21....5M}
{Miko{\l}ajewska} J.,  2012, \mn@doi [Baltic Astronomy]
  {10.1515/astro-2017-0352}, \href
  {https://ui.adsabs.harvard.edu/abs/2012BaltA..21....5M} {21, 5}

\bibitem[\protect\citeauthoryear{{Mikolajewska}, {Acker}  \&
  {Stenholm}}{{Mikolajewska} et~al.}{1997}]{1997A&A...327..191M}
{Mikolajewska} J.,  {Acker} A.,   {Stenholm} B.,  1997, \aap, \href
  {https://ui.adsabs.harvard.edu/abs/1997A&A...327..191M} {327, 191}

\bibitem[\protect\citeauthoryear{{Muerset}, {Nussbaumer}, {Schmid}  \&
  {Vogel}}{{Muerset} et~al.}{1991}]{1991A&A...248..458M}
{Muerset} U.,  {Nussbaumer} H.,  {Schmid} H.~M.,   {Vogel} M.,  1991, \aap,
  \href {https://ui.adsabs.harvard.edu/abs/1991A&A...248..458M} {248, 458}

\bibitem[\protect\citeauthoryear{{Munari} \& {Patat}}{{Munari} \&
  {Patat}}{1993}]{1993A&A...277..195M}
{Munari} U.,  {Patat} F.,  1993, \aap, \href
  {https://ui.adsabs.harvard.edu/abs/1993A&A...277..195M} {277, 195}

\bibitem[\protect\citeauthoryear{{Munari} \& {Zwitter}}{{Munari} \&
  {Zwitter}}{1997}]{1997A&A...318..269M}
{Munari} U.,  {Zwitter} T.,  1997, \aap, \href
  {https://ui.adsabs.harvard.edu/abs/1997A&A...318..269M} {318, 269}

\bibitem[\protect\citeauthoryear{{Murset} \& {Nussbaumer}}{{Murset} \&
  {Nussbaumer}}{1994}]{1994A&A...282..586M}
{Murset} U.,  {Nussbaumer} H.,  1994, \aap, \href
  {https://ui.adsabs.harvard.edu/abs/1994A&A...282..586M} {282, 586}

\bibitem[\protect\citeauthoryear{{Netzer}}{{Netzer}}{1975}]{1975MNRAS.171..395N}
{Netzer} H.,  1975, \mn@doi [\mnras] {10.1093/mnras/171.2.395}, \href
  {https://ui.adsabs.harvard.edu/abs/1975MNRAS.171..395N} {171, 395}

\bibitem[\protect\citeauthoryear{{Neugebauer} et~al.,}{{Neugebauer}
  et~al.}{1984}]{1984ApJ...278L...1N}
{Neugebauer} G.,  et~al., 1984, \mn@doi [\apjl] {10.1086/184209}, \href
  {https://ui.adsabs.harvard.edu/abs/1984ApJ...278L...1N} {278, L1}

\bibitem[\protect\citeauthoryear{{O'Donoghue} et~al.,}{{O'Donoghue}
  et~al.}{2006}]{2006MNRAS.372..151O}
{O'Donoghue} D.,  et~al., 2006, \mn@doi [\mnras]
  {10.1111/j.1365-2966.2006.10834.x}, \href
  {https://ui.adsabs.harvard.edu/abs/2006MNRAS.372..151O} {372, 151}

\bibitem[\protect\citeauthoryear{{Orio}, {Zezas}, {Munari}, {Siviero}  \&
  {Tepedelenlioglu}}{{Orio} et~al.}{2007}]{2007ApJ...661.1105O}
{Orio} M.,  {Zezas} A.,  {Munari} U.,  {Siviero} A.,   {Tepedelenlioglu} E.,
  2007, \mn@doi [\apj] {10.1086/514806}, \href
  {https://ui.adsabs.harvard.edu/abs/2007ApJ...661.1105O} {661, 1105}

\bibitem[\protect\citeauthoryear{{Paczynski} \& {Zytkow}}{{Paczynski} \&
  {Zytkow}}{1978}]{1978ApJ...222..604P}
{Paczynski} B.,  {Zytkow} A.~N.,  1978, \mn@doi [\apj] {10.1086/156176}, \href
  {https://ui.adsabs.harvard.edu/abs/1978ApJ...222..604P} {222, 604}

\bibitem[\protect\citeauthoryear{{Peterson}}{{Peterson}}{1997}]{1997iagn.book.....P}
{Peterson} B.~M.,  1997, {An Introduction to Active Galactic Nuclei}

\bibitem[\protect\citeauthoryear{{Pojmanski}}{{Pojmanski}}{1997}]{1997AcA....47..467P}
{Pojmanski} G.,  1997, \actaa, \href
  {https://ui.adsabs.harvard.edu/abs/1997AcA....47..467P} {47, 467}

\bibitem[\protect\citeauthoryear{{Reinmuth}}{{Reinmuth}}{1925}]{1925AN....225..385R}
{Reinmuth} K.,  1925, \mn@doi [Astronomische Nachrichten]
  {10.1002/asna.19252252302}, \href
  {https://ui.adsabs.harvard.edu/abs/1925AN....225..385R} {225, 385}

\bibitem[\protect\citeauthoryear{{Richwine}, {Bedient}, {Slater}  \&
  {Mattei}}{{Richwine} et~al.}{2005}]{2005JAVSO..34...28R}
{Richwine} P.,  {Bedient} J.,  {Slater} T.,   {Mattei} J.~A.,  2005, Journal of
  the American Association of Variable Star Observers (JAAVSO), \href
  {https://ui.adsabs.harvard.edu/abs/2005JAVSO..34...28R} {34, 28}

\bibitem[\protect\citeauthoryear{{Roberts}, {Lehar}  \& {Dreher}}{{Roberts}
  et~al.}{1987}]{1987AJ.....93..968R}
{Roberts} D.~H.,  {Lehar} J.,   {Dreher} J.~W.,  1987, \mn@doi [\aj]
  {10.1086/114383}, \href
  {https://ui.adsabs.harvard.edu/abs/1987AJ.....93..968R} {93, 968}

\bibitem[\protect\citeauthoryear{{Schmid}}{{Schmid}}{1989}]{1989A&A...211L..31S}
{Schmid} H.~M.,  1989, \aap, \href
  {https://ui.adsabs.harvard.edu/abs/1989A&A...211L..31S} {211, L31}

\bibitem[\protect\citeauthoryear{{Scholz}}{{Scholz}}{1992}]{1992A&A...253..203S}
{Scholz} M.,  1992, \aap, \href
  {https://ui.adsabs.harvard.edu/abs/1992A&A...253..203S} {253, 203}

\bibitem[\protect\citeauthoryear{{Shappee} et~al.,}{{Shappee}
  et~al.}{2014}]{2014AAS...22323603S}
{Shappee} B.,  et~al., 2014, in American Astronomical Society Meeting Abstracts
  \#223. p. 236.03

\bibitem[\protect\citeauthoryear{{Sokoloski}, {Bildsten}  \& {Ho}}{{Sokoloski}
  et~al.}{2001}]{2001MNRAS.326..553S}
{Sokoloski} J.~L.,  {Bildsten} L.,   {Ho} W. C.~G.,  2001, \mn@doi [\mnras]
  {10.1046/j.1365-8711.2001.04582.x}, \href
  {https://ui.adsabs.harvard.edu/abs/2001MNRAS.326..553S} {326, 553}

\bibitem[\protect\citeauthoryear{{Stahl}, {Kaufer}  \& {Tubbesing}}{{Stahl}
  et~al.}{1999}]{1999ASPC..188..331S}
{Stahl} O.,  {Kaufer} A.,   {Tubbesing} S.,  1999, {The FEROS spectrograph}.
p.~331

\bibitem[\protect\citeauthoryear{{Stellingwerf}}{{Stellingwerf}}{1978}]{1978ApJ...224..953S}
{Stellingwerf} R.~F.,  1978, \mn@doi [\apj] {10.1086/156444}, \href
  {https://ui.adsabs.harvard.edu/abs/1978ApJ...224..953S} {224, 953}

\bibitem[\protect\citeauthoryear{{Trabucchi}, {Wood}, {Montalb{\'a}n},
  {Marigo}, {Pastorelli}  \& {Girardi}}{{Trabucchi}
  et~al.}{2019}]{2019MNRAS.482..929T}
{Trabucchi} M.,  {Wood} P.~R.,  {Montalb{\'a}n} J.,  {Marigo} P.,  {Pastorelli}
  G.,   {Girardi} L.,  2019, \mn@doi [\mnras] {10.1093/mnras/sty2745}, \href
  {https://ui.adsabs.harvard.edu/abs/2019MNRAS.482..929T} {482, 929}

\bibitem[\protect\citeauthoryear{{Vassiliadis} \& {Wood}}{{Vassiliadis} \&
  {Wood}}{1993}]{1993ApJ...413..641V}
{Vassiliadis} E.,  {Wood} P.~R.,  1993, \mn@doi [\apj] {10.1086/173033}, \href
  {https://ui.adsabs.harvard.edu/abs/1993ApJ...413..641V} {413, 641}

\bibitem[\protect\citeauthoryear{{Webster} \& {Allen}}{{Webster} \&
  {Allen}}{1975}]{1975MNRAS.171..171W}
{Webster} B.~L.,  {Allen} D.~A.,  1975, \mn@doi [\mnras]
  {10.1093/mnras/171.1.171}, \href
  {https://ui.adsabs.harvard.edu/abs/1975MNRAS.171..171W} {171, 171}

\bibitem[\protect\citeauthoryear{{Whitelock}, {Menzies}, {Feast}, {Marang},
  {Carter}, {Roberts}, {Catchpole}  \& {Chapman}}{{Whitelock}
  et~al.}{1994}]{1994MNRAS.267..711W}
{Whitelock} P.,  {Menzies} J.,  {Feast} M.,  {Marang} F.,  {Carter} B.,
  {Roberts} G.,  {Catchpole} R.,   {Chapman} J.,  1994, \mn@doi [\mnras]
  {10.1093/mnras/267.3.711}, \href
  {https://ui.adsabs.harvard.edu/abs/1994MNRAS.267..711W} {267, 711}

\bibitem[\protect\citeauthoryear{{Whitelock}, {Marang}  \& {Feast}}{{Whitelock}
  et~al.}{2000}]{2000MNRAS.319..728W}
{Whitelock} P.,  {Marang} F.,   {Feast} M.,  2000, \mn@doi [\mnras]
  {10.1046/j.1365-8711.2000.03743.x}, \href
  {https://ui.adsabs.harvard.edu/abs/2000MNRAS.319..728W} {319, 728}

\bibitem[\protect\citeauthoryear{{Whitelock}, {Feast}, {van Loon}  \&
  {Zijlstra}}{{Whitelock} et~al.}{2003}]{2003MNRAS.342...86W}
{Whitelock} P.~A.,  {Feast} M.~W.,  {van Loon} J.~T.,   {Zijlstra} A.~A.,
  2003, \mn@doi [\mnras] {10.1046/j.1365-8711.2003.06514.x}, \href
  {https://ui.adsabs.harvard.edu/abs/2003MNRAS.342...86W} {342, 86}

\bibitem[\protect\citeauthoryear{{Whitelock}, {Feast}  \& {Van
  Leeuwen}}{{Whitelock} et~al.}{2008}]{2008MNRAS.386..313W}
{Whitelock} P.~A.,  {Feast} M.~W.,   {Van Leeuwen} F.,  2008, \mn@doi [\mnras]
  {10.1111/j.1365-2966.2008.13032.x}, \href
  {https://ui.adsabs.harvard.edu/abs/2008MNRAS.386..313W} {386, 313}

\bibitem[\protect\citeauthoryear{{Woitke}}{{Woitke}}{2006}]{2006A&A...460L...9W}
{Woitke} P.,  2006, \mn@doi [\aap] {10.1051/0004-6361:20066322}, \href
  {https://ui.adsabs.harvard.edu/abs/2006A&A...460L...9W} {460, L9}

\bibitem[\protect\citeauthoryear{{Zamanov} et~al.,}{{Zamanov}
  et~al.}{2017}]{2017AN....338..680Z}
{Zamanov} R.~K.,  et~al., 2017, \mn@doi [Astronomische Nachrichten]
  {10.1002/asna.201713362}, \href
  {https://ui.adsabs.harvard.edu/abs/2017AN....338..680Z} {338, 680}

\makeatother
\end{thebibliography}

%%%%%%%%%%%%%%%%%%%%%%%%%%%%%%%%%%%%%%%%%%%%%%%%%%

% Don't change these lines
\bsp	% typesetting comment
\label{lastpage}
\end{document}